\documentstyle[12pt,epsf]{article}
\def\boxit#1{\kern4pt\vbox{\hrule\hbox{\vrule\kern8pt\vbox{\kern8pt#1\kern8pt}\ern8pt\vrule}\hrule}}

\def\and{ {\rm and} }

\begin{document}
\onecolumn

\vfill

\begin{center}
{\Large \bf 
Test of the Equivalence Principle from K Physics} \\ \vfill
%Violation of equivalence principle in the K-system} \\ \vfill
T. Hambye$^{(a)}$\footnotemark\footnotetext{email: 
hambye@hal1.physik.uni-dortmund.de}, 
R.B. Mann$^{(b)}$\footnotemark\footnotetext{email: 
mann@avatar.uwaterloo.ca}
and U. Sarkar$^{(c)}$\footnotemark\footnotetext{email: utpal@prl.ernet.in}\\
\vspace{2cm}
(a) Institut f\"{u}r Physik, Universit\"{a}t Dortmund,
D-44221 Dortmund, Germany \\
(b) Physics Department, University of Waterloo,
Waterloo, Ontario, Canada N2L 3G1\\
(c)Theory Group, Physical Research Laboratory,
Ahmedabad, 380 009, India\\
\vspace{2cm}
\end{center}

\vfill

\begin{abstract}
A violation of the equivalence principle (VEP) in the Kaon system
can, in principle, induce
oscillations between $K^\circ$ and $\overline{K^\circ}$
in a manner that need not violate CPT conservation. 
We show that such a CPT-conserved VEP mechanism could
be clearly tested experimentally through the energy dependence of 
the $K_L-K_S$ mass difference and discuss constraints imposed
by present experiments. 
\end{abstract}

\vfill
%\clearpage

%\twocolumn

The  principle  of  equivalence  (EP)  implies   universality  of
gravitational  coupling  for all  forms of  mass-energy,  thereby
ensuring  that  spacetime is  described  by a unique  operational
geometry.  Although the EP has been tested to  impressive  levels
of precision, virtually all such tests have been carried out with
matter  fields.  The  possibility  that matter and antimatter may
have different gravitational couplings remains a fascinating open
question.  The strongest bound on matter-antimatter gravitational
universality  comes  from  the  $K^\circ  -   \overline{K^\circ}$
system.  However all studies of this system have considered
a straightforward violation of the weak equivalence principle (WEP) 
in which it is assumed that $K^\circ  -   \overline{K^\circ}$
mass and gravitational eigenstates can be simultaneously diagonalised 
but with differing eigenvalues ({\it i.e.} differing 
$K^\circ$ and $\overline{K^\circ}$ masses)
\cite{good,Kenyon,hughes}, in which case violation of  gravitational
universality also means violation of CPT.

More generally, a violation of the EP (VEP) in the Kaon system will 
not assume simultaneous pairwise diagonalization of
mass, gravitational or weak eigenstates.
We shall consider in
this letter the consequences of such a general VEP mechanism, showing
that it can provide a new source of CP--violation whilst conserving
CPT. In this context, the 
WEP violation investigated previously \cite{good,Kenyon,hughes}
is a special case of maximal CPT violation in the gravitational sector.
We consider constraints imposed on the VEP mechanism by present 
experiments; constraints on VEP-induced CP--violation
will be discussed elsewhere \cite{HMS}.

Since gravitational couplings are proportional to  mass-energy one needs 
to study the energy of the particles under consideration and to consider
relativistic kaons for which  gravitational effects could be 
non-negligible. In a weak gravitational
field $g_{\mu\nu}=\eta_{\mu\nu}+ h_{\mu\nu}$ (where $h_{\mu\nu}=
2\frac{\phi}{c^2} {\mbox diag}(1,1,1,1)$) the gravitational part of
the Lagrangian is $-\frac{1}{2}(1+g_i)h_{\mu\nu}T^{\mu\nu}$ 
where $T^{\mu\nu}$ is the stress-energy in the gravitational eigenbasis
and $1+g_i$ are the gravitational couplings,
which are not equal if VEP is operative. For relativistic 
pointlike kaons
the general form of the effective Hamiltonian in the
$(K^\circ \hskip .15in \overline{K^\circ})$ basis will be 
\begin{equation}
H = p  I + U_W H_{SEW} U_W^{-1} + U_G H_G U_G^{-1} \label{h}
\end{equation}
with I the identity matrix, 
\begin{equation}
H_{SEW} = \frac{(M_{SEW})^2}{2 p} = \frac{1}{2 p} {\pmatrix{ 
m_1 & 0 \cr 0 & m_2}}^2 \label{hsew}
\end{equation}
and
\begin{equation}
H_G = \pmatrix{ G_1 & 0 \cr 0 & G_2} = \pmatrix{ 
- 2 (1 + g_1) \phi (p + \frac{\bar{m}^2}{2 p}) & 0 \cr 0 &
- 2 (1 + g_2) \phi (p + \frac{\bar{m}^2}{2 p})} \label{hg}
\end{equation}
in physical time and length units \cite{hughes}
to first order in $\bar{m}^2/p^2$ with p the momentum
and $\bar{m}$, where for a quantity $X$,
$\delta X \equiv(X_1-X_2)$, $\bar{X} = (X_1+X_2)/2$.
 $H_{SEW}$ is the 
matrix coming from the  strong, electromagnetic and
weak interactions, whose absorptive ({\it i.e.} antihermitian)
parts we shall neglect for the moment. In the absence of gravity,
weak interactions are responsible for $m_1 \not= m_2$, which
are interpreted as the $K_L$ and $K_S$ masses.
We have neglected terms proportional to $\delta{m} \equiv
(m_1 - m_2)/2$ in the gravitational Hamiltonian
$H_G$. $\phi$ is the 
gravitational potential on the surface of earth, which is constant over
the range of terrestrial experiments. 

Since  $H_{SEW}$ and $H_G$ are hermitian, then $U_G$ and $U_W$ are unitary.
{}From the general form of a 2x2 unitary matrix
$$
U = e^{i\chi}
\pmatrix{ e^{-i\alpha} & 0 \cr 0 & e^{i\alpha}} 
\pmatrix{ \cos\theta & \sin\theta \cr -\sin\theta & \cos\theta} 
\pmatrix{ e^{i\beta} & 0 \cr 0 & e^{i\beta}} 
$$
it is straightforward to show that
$$
H = p I + \frac{1}{2 p} { \pmatrix{ M_{+} & M_{12} \cr 
M_{12}^\ast & M_{-} }}^2
$$
with
\begin{eqnarray}
M_{\pm} &=& \bar{m} + \frac{p}{\bar{m}} \bar{G} \pm \frac{cos2\theta_W}{2}
\delta m \pm \frac{p}{\bar{m}} \frac
{\cos2\theta_G}{2}\delta G \nonumber \\
M_{12} &=& -(\sin2\theta_W \delta m 
+ e^{-2i(\alpha_G-\alpha_W)}\frac{p}{\bar{m}}\sin2\theta_G \delta G)/2 \nonumber \\
\end{eqnarray} \label{mtot}
where we have absorbed additional phases into the
$K^\circ$ and $\overline{K^\circ}$ wavefunctions. Note
that gravitation may induce CP--violation
if $\alpha_G \neq \alpha_W$; we shall study the implications of this
elsewhere. Since in this paper we will be considering effects for which
CP-violation is negligible, for simplicity we shall
take $\alpha_G$ =$\alpha_W$ = 0.

In the basis of the 
physical states $K_L$ and $K_S$, the Hamiltonian becomes
\begin{equation}
H = \pmatrix{p + \frac{m_L^2}{2 p} & 0 \cr 0 & p + \frac{m_S^2}{2 p} }
= \pmatrix{\tilde{E} & 0 \cr 0 & 
 \tilde{E}} + {1 \over 2}
\pmatrix{ \Delta E & 0 \cr 0 & - \Delta E } \label{he}
\end{equation}
where $\tilde{E} = (1 - 2 (1 + \bar{g})\phi) (p + \frac{\bar{m}^2}{2 p} )$,
and

\begin{equation}
\frac{p}{\bar{m}} \Delta E = m_L - m_S = \left[ 
( \delta m )^2 + {\left( 2 \phi \delta g {p \over \bar{m}}
(p + {\bar{m}^2 \over 2 p}) \right)}^2 - 4 \delta m \phi \delta g
\frac{p}{\bar{m}} (p + {\bar{m}^2 \over 2 p} ) 
\cos(2 \theta_W - 2 \theta_G) \right]^{1 / 2} 
\label{mls}
\end{equation}
where $m_L$ and $m_S$ are the experimentally measured masses of 
$K_L$ and $K_S$ respectively.

%Eq.(\ref{mls}) will be the relevant equation for further studies together
%with the following equation which gives 
The amount of CPT--violation is given by
\begin{equation}
\Delta_{CPT} = M_{+} - M_{-} = \cos (2 \theta_W)\delta m - 
\cos (2  \theta_G) 2 \phi \delta g \frac{p}
{\bar{m}} (p + {\bar{m}^2 \over 2 p}) \label{dcpt} 
\end{equation}
Previous studies of VEP in the Kaon system assumed CPT violation
in the gravitational sector only, from which  
 it was argued that empirical bounds 
can be placed on the difference between the gravitational couplings
to $|K^\circ>$ and $|\overline{K^\circ}>$. 
The difference in gravitational 
eigenvalues then corresponds to a difference ($\Delta M_g$) 
in the masses of  $|K^\circ>$ and $|\overline{K^\circ}>$:
\begin{equation}
\vert M_{+} - M_{-}\vert = \phi \Delta M_g = 2
\phi \vert\delta g\vert \frac{p}{\bar{m}} (p + {\bar{m}^2 \over 2 p}) \label{kh}
\end{equation}
and is entirely attributable to the
amount of CPT violation. The first equality in Eq.(\ref{kh}) 
was given by  Kenyon \cite{Kenyon} and the second by
Hughes \cite{hughes}, who specified the energy dependence 
of $\Delta M_g$. From the experimental upper bound on $M_{+} - M_{-}$ 
\cite{Carosi} the bound $\mid\delta g\mid  < 2.5\times10^{-18}$ may
be obtained, where the potential $\phi$ is taken to be that due to
the local supercluster ($\phi \simeq 3\times 10^{-5}$).
In this approach CPT conservation implies
no gravitational mass difference and hence no VEP.
However it is clear from the expression (\ref{dcpt})
for $\Delta_{CPT}$ that the bound obtained 
on $\Delta M_g$ is actually on some combination of 
VEP parameters and
not on $\delta g$ and $\cos (2 \theta_G)$ separately.
When $\theta_G = 0$, $\theta_M = \pi/4$, Eq.(\ref{kh}) 
agrees with Eq.(\ref{dcpt}). More recent experiments \cite{pdg} find
$|M_{+}-M_{-}|/m_K < 9\times 10^{-19}$, yielding the bound
$\mid\delta g\mid < 4.5\times 10^{-19}$ for the same 
value  of $\phi$.

Next we shall consider a scenario in which CPT is conserved,
so that $\Delta_{CPT} = 0$. From the above it is clear that, even if CPT
is conserved,
there is still a VEP-induced difference between the masses of 
the physical states.  As a result bounds can be placed
on the VEP parameter $\phi\delta g$ without the assumption that
locality in quantum field theory is violated. One 
interesting consequence of this is that
the VEP mechanism predicts that the mass difference $m_L - m_S$
will be energy dependent.

{}From the expression of $\Delta_{CPT} $ it is clear that it is not 
possible to conserve CPT for all momenta unless
$\theta_M = \theta_G = {\pi \over 4}$ (modulo $\pi$), thereby
separately conserving CPT in the weak and the gravitational 
sectors. In this case the mass difference is 
\begin{equation}
m_L - m_S =  \delta m - 2 \phi \delta g {{p} \over
 {\bar{m}}} (p + {\bar{m}^2 \over 2 p}) .
\label{bef}
\end{equation} 
which as noted above is energy dependent.
It is possible to put a bound on the VEP parameter
$\delta g$ if we know the value of $\phi$ and 
the mass  difference at various given energies. Alternatively, if mass 
measurements at two different energies were different, the
differing values for $m_L - m_S$ could be used to
extract a value for the VEP parameter $\delta g$.

What constraints do present experiments place 
on $\delta m$ and $\delta g$? 
In the review of particle properties \cite{pdg} six experiments were taken
into account. Two of them are at high energy \cite{gib,sch} 
with the kaon momentum $p_K$
between 20 GeV and 160 GeV. The weighted average of these two 
experiments is \cite{sch}: $\Delta m_{LS} = m_L - m_S
= (0.5282 \pm 0.0030) 10^{10} \hbar s^{-1}$. 
The four other experiments \cite{cullen,gew,gjes,adler} are
at lower energy, with $p_K \approx 5$ GeV, or less. The weighted average
of these low energy experiments
is $\Delta m_{LS} = (0.5322 \pm 
0.0018) 10^{10} \hbar s^{-1}$.
A fit of equation (\ref{bef}) with the high and low energy value of 
$\Delta m_{LS}$ gives : $\delta m = (3.503 \pm 0.012) \times
10^{-12} MeV$ and 
$\phi \delta g = (8.0 \pm 7.0) \times 10^{-22} \times \left( \frac{90}{E_{av}}
\right)^2$, (where $E_{av}$ is the average energy for the high energy
experiment). 

Taking $\phi$ to be the earth's potential ($\phi \simeq 0.69\times
10^{-9}$), 
we find $\delta g = (1.2 \pm 1.0)\times 10^{-12}$ whereas 
if $\phi$ is due to the local supercluster then 
$\delta g = (2.7 \pm 2.3)\times 10^{-17}$.
These values differ from zero by 1.15 standard 
deviations. While it is certainly premature to regard this
as evidence for the VEP mechanism, it does 
show that VEP in the Kaon sector is not as tightly constrained as
previous studies \cite{Kenyon,hughes} have implied.
A precise fit of mass difference
per energy bin in present and future high energy
experiments would be extremely useful in constraining 
the VEP parameters, particularly since the
present experimental situation at low energy is not clear. Indeed
one of the low energy experiments \cite{adler} was published last year
found $\Delta m_{LS} 
= (0.5274 \pm 0.0029 \pm 0.0005) 10^{10} \hbar s^{-1}$, a value
lower than the weighted average 
$\Delta m_{LS} = (0.5350 \pm 0.0023) 10^{10} \hbar s^{-1}$
of the three (previous) low energy 
experiments. 
Without this new experiment, a similar fit of the other five experiments
yields
$\phi \delta g = (1.38 \pm 0.77) \times 10^{-21} (90/E_{av})^2$. In 
this case $\delta g$ is different from 0 by 1.8 standard deviations. 
Alternatively taking only the new experiment \cite{adler} at 
low energy we would obtain a value compatible with 0 at less 
than 1 standard deviation.

In the above analysis we have not included the effect of the absorptive
part of the Hamiltonian, which if VEP is operative appears
in both the weak and gravitational sectors.
Here we consider the absorptive part coming from the weak sector.

In the weak sector, inclusion of the 
absorbtive part entails the replacement $m_i$ by $m_i - i
 \Gamma_i/2$. With this change the definitions of $\tilde{E}$
and $\Delta E$ are modified to
\begin{eqnarray}
\tilde{E} &= & \left( p + \displaystyle \frac{(\bar{m} - 
i \bar{\Gamma}/2)^2}{2 p} \right) (1 - 2 (1 + \bar{g}) \phi) \nonumber \\
\frac{p}{\bar{m}} \Delta E &=& {1 \over
 \sqrt{2}} \left[ \sqrt{F^2 + G^2} + F \right]^{1/2} 
+ i {1 \over \sqrt{2}} \left[ \sqrt{F^2 + G^2} - F \right]^{1/2} 
\nonumber \\
F & = & (\delta m)^2 + (2 \phi \delta g {{p} \over {\bar{m}}}(p + 
{\bar{m}^2 \over 2 p} 
))^2 - 4 
\delta m \phi \delta g {{p} \over
 {\bar{m}}} (p + {\bar{m}^2 \over 2 p}) \cos (2 \theta_W - 2 \theta_G)
- ( {{\delta \Gamma} \over 2})^2 \nonumber \\
G &=& - ( \delta m  \delta \Gamma) + 2 \cos (2 \theta_W -
2 \theta_G) [ \delta \Gamma \phi \delta g {{p} \over {\bar{m}}} (p +
 {\bar{m}^2 \over 2p}) ]
\end{eqnarray}
We also have,
\begin{eqnarray}
m_L - m_S &= & \frac{p}{\bar{m}} {\rm Re} (\Delta E) \label{ls} \\
\Gamma_S - \Gamma_L &=&  2 \frac{p}{\bar{m}} {\rm Im} (\Delta E) \label{ps}
\end{eqnarray}
In deriving these equations we neglected terms in $\delta m  \Gamma$,
$\delta m \delta \Gamma$ and $\Gamma^2$ with respect to the terms in
$m \delta m$ or $m \delta \Gamma$. 
It can be shown that in the $CPT$--conserving case the above mass
difference (equation (\ref{ls})) reduces to equation (\ref{bef}).
So in the CPT conserving case the results above are not affected
by inclusion of the widths. In this case the difference
$\Gamma_S - \Gamma_L = \delta \Gamma$ is independent of energy.
This is consistent with experiment, which indicates that
the low and high energy measurements of $\Gamma_S - \Gamma_L$ are
fully compatible \cite{pdg}. 
For $\theta_G \not= \pi/4$, an examination of (\ref{ps})
indicates that $\Gamma_S - \Gamma_L$ is energy dependent; however 
this is small and measurements of $\Gamma_S - \Gamma_L$ do not
constrain $\delta g$ more than measurements of $\Delta m_{LS}$ even
though they are more precise. We note that
measurements of $\Gamma_S - \Gamma_L$ would more strongly constrain
a possible absorptive part coming from the gravitational sector
which presumably would induce a larger energy dependence.
We shall not consider this possibility here.
For $\theta_G \not= \pi/4$, width effects
in Eq.(\ref{ls}) are small and Eq.(\ref{mls}) remain 
good up to a few percents.

In Fig.1, for completeness, we plot as a function of $\cos (2\theta_G)$ (and
with $\theta_W = \pi/4$) the upper bounds we get 
on $\vert \phi \delta g\vert$ by fixing
$\delta m$ to the central value of the world average \cite{pdg},
$\Delta m_{LS}$ =$(0.5310 \pm 0.0019) 10^{10} \hbar s^{-1}$ and 
requiring that
$m_L -m_S$ in (\ref{mls}) does not differ from $\delta m$ by more that
$\pm 2$ standard deviations. Note that in the case of maximal CPT 
violation ($\theta_G = 0$), $m_L-m_S$ can only increase with energy,
as is clear from Eq.(\ref{mls}) or Eq. (\ref{ls}). The actual difference
between low and high energy experiments, if valid, 
could not be explained in this case except for complex value of 
$\delta g$ (and similarly for values
of $\theta_G$ very close to 0). In Fig.1 we also show the bound coming
from Eq.(\ref{dcpt}) with $\theta_W = \pi/4$ and
$|M_{+}-M_{-}|/m_K < 9\times 10^{-19}$ \cite{pdg}. These curves have
different interpretations. The solid line is obtained
from constraints on possible energy dependence of $\delta m_{LS}$,
a distinctive signal of VEP. The dashed line is obtained from
constraints on CPT violation effects. These need not originate 
from the VEP mechanism, and could be relaxed if other (energy independent)
CPT violating effects were operative. As is clear from Eq. (\ref{mls}),
additional CPT violation effects coming from $\theta_W \neq 0$ would
not qualitatively change these bounds. The solid line is therefore a more
secure limit for $\theta_G \neq 0$.

Violations of the equivalence principle in the Kaon system need not
violate CPT (which in turn implies a loss of locality in quantum 
field theory).   More precise and detailed tests in this sector
should provide us with important empirical information on the
validity of the equivalence principle.

\vspace{1cm}

\noindent
{\bf Acknowledgements}

This work was supported in part by the Natural Sciences and
Engineering Research Council of Canada. One of us (US) acknowledges 
hospitality at the Institut fur Physik, Univ Dortmund, Germany and
a fellowship from the Alexander von Humboldt Foundation.

Fig.1: Upper bounds on $\vert \phi \delta g\vert$ (for p $\simeq 90 GeV$)
obtained from Eq.(\ref{mls}) (solid line) and Eq.(\ref{dcpt})(dashed line)
as explained in the text.

\begin{thebibliography}{99}
\baselineskip 16pt
\bibitem{good} M.L. Good, Phys. Rev. {\bf 121}, 311 (1961).
\bibitem{Kenyon} I.R. Kenyon, Phys. Lett. {\bf B237}, 274 (1990).
\bibitem{hughes} R.J. Hughes, Phys. Rev. {\bf D46}, R2283 (1992);
\bibitem{HMS}T. Hambye, R.B. Mann and U. Sarkar, in preparation.
\bibitem{Carosi}R. Carosi {\it et.al.} Phys. Lett. {\bf B237}, 303 (1990).
\bibitem{pdg}Review of Particle Properties, Phys. Rev. {\bf D54}, 1 (1996).
\bibitem{gib} L.K. Gibbons, et al., Phys. Rev. Lett. {\bf 70}, 1199 (1993).
\bibitem{sch} B. Schwingenheuer et al., Phys. Rev. Lett. {\bf 74}, 4376
 (1995).
\bibitem{cullen} M. Cullen, et al., Phys Lett. {\bf 32 B}, 523 (1970).
\bibitem{gew} C.Geweniger et al., Phys. Lett. {\bf 48 B}, 487 (1974).
\bibitem{gjes} Gjesdal et al., Phys. Lett. {\bf 52 B}, 113 (1974).
\bibitem{adler} R. Adler et al., Phys. Lett. {\bf B 363}, 237 (1995).

%\bibitem{barmin} V.V. Barmin et al., Nucl. Phys. {\bf B 247}, 293 (1984).
%\bibitem{barish}B.C. Barish, Nucl. Phys. {\bf B38} (Proc. Supp.), 343 (1995). 
%\bibitem{aron}S.H. Aronson, G.J. Bock, H-Y Cheng and E. Fishbach,
% {\bf 48} (1982) 1306; Phys. Rev. {\bf D28}, 495 (1983).

\end{thebibliography}
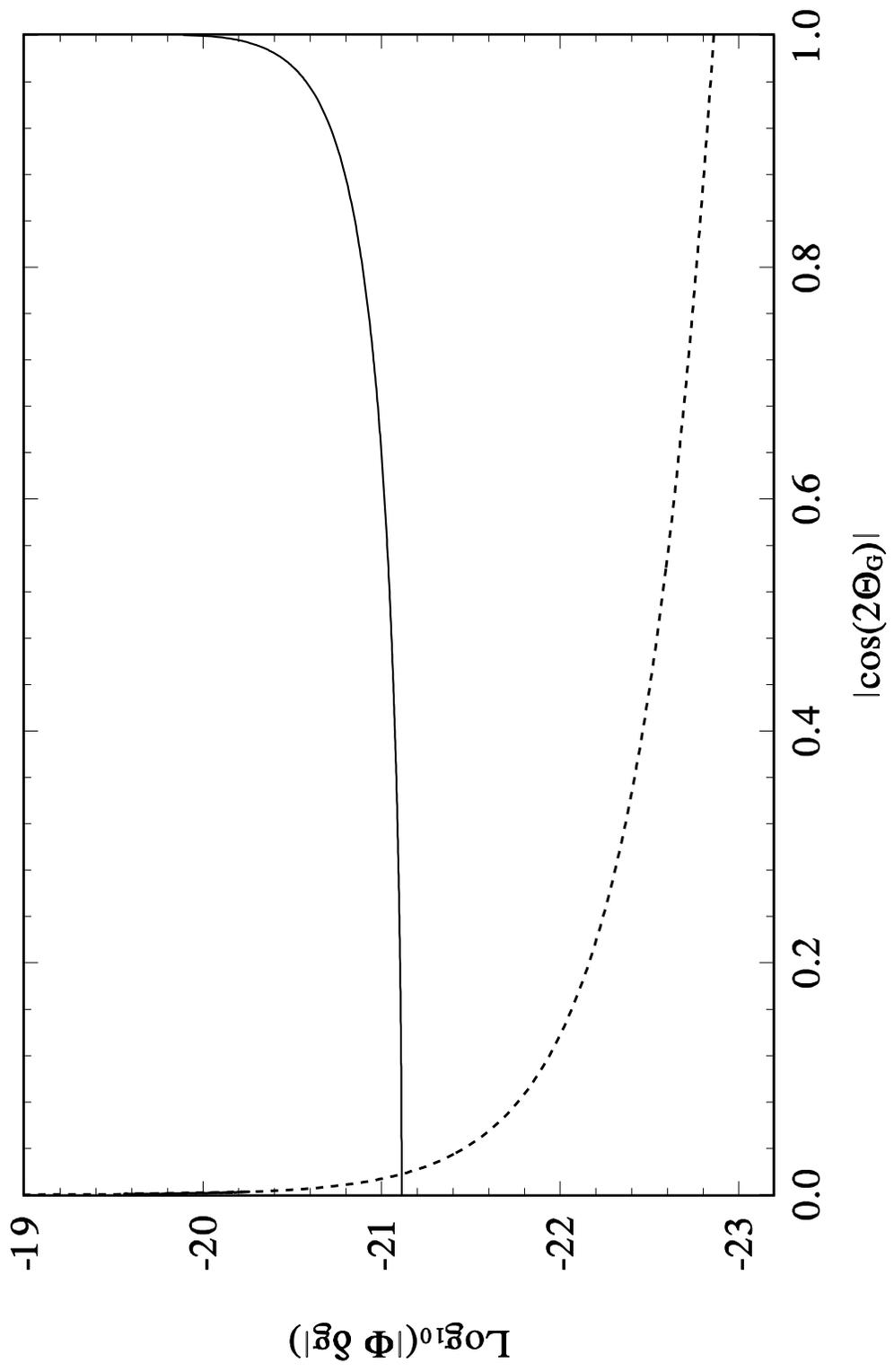
\end{document}